\documentclass[prl,reprint,onecolumn,amsmath,amssymb,floatfix,footinbib,superscriptaddress,longbibliography,openany]{revtex4}
\usepackage{dcolumn}
\usepackage{graphicx}
\usepackage{mathrsfs}
\usepackage{mdwlist}
\usepackage{subfigure}
\usepackage{booktabs}
\usepackage{amsmath}
\usepackage{dsfont}
\usepackage{amstext}
\usepackage{amssymb}
\usepackage{amsbsy}
\usepackage{bbm}
\usepackage{amsthm}
\usepackage{graphicx}
\usepackage{textcomp}
\usepackage{multirow}
\usepackage{color}
\usepackage[colorlinks,citecolor=blue]{hyperref}
\definecolor{Dgreen}{RGB}{0, 100, 0}
\usepackage{url}
\usepackage[colorlinks]{hyperref}
\hypersetup{%
	plainpages=true,
	breaklinks=true,  
	hypertexnames=false,  
	pageanchor=true,
	colorlinks=true,
	linkcolor={blue},
	citecolor={red},
	urlcolor={blue},
	anchorcolor={black}
}

\hypersetup{%
	plainpages=true,
	breaklinks=true,  
	hypertexnames=false,  
	pageanchor=true,
	colorlinks=true,
	linkcolor={blue},
	citecolor={red},
	urlcolor={blue},
	anchorcolor={black}
}

 \makeatletter

\newcommand{\Rmnum}[1]{\expandafter\@slowromancap\romannumeral #1@}
\makeatother

\usepackage{CJKutf8}

\begin{document}
	
	\title{Sudden change of the photon output field marks phase transitions in the
		quantum Rabi model}%
	
	\author{Ye-Hong Chen}\thanks{yehong.chen@fzu.edu.cn}
	\affiliation{Fujian Key Laboratory of Quantum Information and Quantum Optics, Fuzhou University, Fuzhou 350116, China}
	\affiliation{Department of Physics, Fuzhou University, Fuzhou 350116, China}
	\affiliation{Theoretical Quantum Physics Laboratory, RIKEN Cluster for Pioneering Research, Wako-shi, Saitama 351-0198, Japan}
	
	\author{Yuan Qiu}%
	\affiliation{Fujian Key Laboratory of Quantum Information and Quantum Optics, Fuzhou University, Fuzhou 350116, China}
	\affiliation{Department of Physics, Fuzhou University, Fuzhou 350116, China}

	\author{Adam Miranowicz}%
	\affiliation{Theoretical Quantum Physics Laboratory, RIKEN Cluster for Pioneering Research, Wako-shi, Saitama 351-0198, Japan}%
	\affiliation{Institute of Spintronics and Quantum Information, Faculty of Physics, Adam Mickiewicz University, 61-614 Pozna\'{n}, Poland}%
	
	\author{Neill Lambert}%
	\affiliation{Theoretical Quantum Physics Laboratory, RIKEN Cluster for Pioneering Research, Wako-shi, Saitama 351-0198, Japan}%
	
	\author{Wei Qin}%
	\affiliation{Theoretical Quantum Physics Laboratory, RIKEN Cluster for Pioneering Research, Wako-shi, Saitama 351-0198, Japan}%
	
	\author{Roberto Stassi}%
	\affiliation{Theoretical Quantum Physics Laboratory, RIKEN Cluster for Pioneering Research, Wako-shi, Saitama 351-0198, Japan}%
	\affiliation{Dipartimento di Scienze Matematiche e Informatiche, Scienze Fisiche e Scienze della Terra, Universit\`{a} di Messina, 98166, Messina, Italy}%
	
	\author{Yan Xia}\thanks{xia-208@163.com}%
	\affiliation{Fujian Key Laboratory of Quantum Information and Quantum Optics, Fuzhou University, Fuzhou 350116, China}%
	\affiliation{Department of Physics, Fuzhou University, Fuzhou 350116, China}%
	
	\author{Shi-Biao Zheng}%
	\affiliation{Fujian Key Laboratory of Quantum Information and Quantum Optics, Fuzhou University, Fuzhou 350116, China}%
	\affiliation{Department of Physics, Fuzhou University, Fuzhou 350116, China}%
	
	\author{Franco Nori} \thanks{fnori@riken.jp}%
	\affiliation{Theoretical Quantum Physics Laboratory, RIKEN Cluster for Pioneering Research, Wako-shi, Saitama 351-0198, Japan}%
	\affiliation{Quantum Information Physics Theory Research Team, RIKEN Center for Quantum Computing, Wako-shi, Saitama 351-0198, Japan}%
	\affiliation{Department of Physics, University of Michigan, Ann Arbor, Michigan 48109-1040, USA}%

\begin{abstract}
\textbf{Abstract:} The experimental observation of quantum phase transitions predicted by the quantum Rabi model in quantum critical systems is usually challenging due to the lack of signature experimental observables associated with them. Here, we describe a method to identify the dynamical critical phenomenon in the quantum Rabi model consisting of a three-level atom and a cavity at the quantum phase transition. Such a critical phenomenon manifests itself as a sudden change of steady-state output photons in the system driven by two classical fields, when both the atom and the cavity are initially unexcited. The process occurs as the high-frequency pump field is converted into the low-frequency Stokes field and multiple cavity photons in the normal phase, while this conversion cannot occur in the superradiant phase. The sudden change of steady-state output photons is an experimentally accessible measure to probe quantum phase transitions, as it does not require preparing the equilibrium state.
\end{abstract}
	\date{\today}

\maketitle

\section*{\large{Introduction}}

In quantum systems close to critical points, small variations of
physical parameters can lead to drastic changes in the
equilibrium-state properties \cite{Agarwal2012Book,Scully1997Book,Scully2009Sci,Sachdev2011book}.
An interesting class of quantum critical
systems is provided by light-matter interaction models \cite{Cong2016,Kirton2018AQT,Kockum2019NRP,Forn2019RMP},
such as the quantum Rabi \cite{Ashhab2013PRA,Hwang2015PRL} and Dicke models \cite{Hepp1973AP,Wang1973PRA,EmaryPRL2003,Lambert2004PRL,Scully2015PRL,Shammah2017PRA,Shammah2018Pra,Makihara2021NC} describing the interaction of single or many two-level atoms (atomic levels $|g\rangle$ and $|e\rangle$) with a single-model cavity.
The quantum Dicke model exhibits a superradiant quantum phase transition (QPT) in the thermodynamic limit of infinite atoms \cite{Hepp1973AP,Wang1973PRA,EmaryPRL2003,Lambert2004PRL,Scully2015PRL,Shammah2017PRA,Shammah2018Pra}.
Such a thermodynamic limit leads to some difficulties in experimentally
exploring the superradiant QPTs \cite{Baumann2010Nat,Baumann2011PRL,Bastidas2012PRL,Mlynek2014NC,Fuchs2016JPB,Zhang2017Opt,Lonard2017Nat,Zhu2020PRL}.
Instead, by replacing the thermodynamic
limit with a scaling of the system parameters, {the quantum Rabi model described by the Hamiltonian (hereafter $\hbar=1$)}
\begin{align}\label{eq1}
	H_{R}=\omega a^{\dag}a(|e\rangle\langle e| +|g\rangle\langle g|)+{\Omega}|e\rangle\langle e|
	-g(a+a^{\dag})(|g\rangle\langle e|+|e\rangle\langle g|),
\end{align}
can also exhibit 
a superradiant QPT \cite{Ashhab2013PRA,Hwang2015PRL}, where $\omega$ ($\Omega$) is the cavity-mode (atomic-transition) frequency, $a$ ($a^{\dag}$) is the cavity-mode 
annihilation (creation) operator, and $g$ is the light-matter coupling strength.
We assume that the level frequency of the atomic state $|g\rangle$ is zero. This superradiant QPT has been experimentally observed \cite{Cai2021NC,Chen2021NC} and is attracting
increasing attention \cite{Puebla2016PRA,Shen2017PRA,Liu2017PRL,Hwang2018PRA,Jiang2021PRA,Zheng2022arXiv}.  
{However, one of the most important critical phenomena in this system, i.e., 
a discontinuity of the derivative for the number of bare photons at the critical point, is hard to observe.}

Unlike classical phase transitions, superradiant QPTs can occur when changing
the system parameters at zero temperature \cite{Cong2016,Kirton2018AQT,Shammah2017PRA}.
Specifically, when $g$ approaches the critical point, it was predicted \cite{Hwang2015PRL} that 
the mean photon number in the ground eigenstate of $H_{R}$ 
suddenly increases to infinity.
This corresponds to a QPT from a normal phase (NP), where
the ground state of the cavity field is not occupied, to a superradiant phase (SP), where the ground state is macroscopically
occupied.
However, experimentally exploring this critical phenomenon is challenging
because: (i) 
the time required to prepare this equilibrium state diverges \cite{Hwang2015PRL};
and (ii) these photons are virtual, so cannot be directly measured \cite{Forn2019RMP,Kockum2019NRP}.
Especially, the difficulty (i) may make transition-edge sensing protocols \cite{Garbe2020PRL,Tsang2013Pra,Wang2014Njp} not effective
in detecting this kind of phase transitions.

Here we show how to overcome these problems by introducing 
additional low-energy levels in the quantum Rabi model and driving transitions among these levels.
We show that the stimulated emission effect of the whole system 
can directly reflect the change of the photon-number distributions of the quantum Rabi Hamiltonian.
The process can be interpreted as a multi-photon down-conversion induced by stimulated Raman transitions [i.e., a pump photon is converted into a Stokes photon and multiple cavity photons, as shown in Fig.~1(a)] \cite{Stassi2013PRL,Huang2014PRA,Kockum2017PRA,Chen2021PRR}.
This parametric down-conversion process changes abruptly when the superradiant QPT occurs in the quantum Rabi Hamiltonian.
Such a change can be observed by measuring the real photons continuously released from the cavity. 
Note that this parametric down-conversion process was initially studied by Huang \emph{et al.}
in 2014 for photon emission via vacuum-dressed intermediate states \cite{Huang2014PRA}.
We find that such a photon emission can be useful in observing quantum critical phenomena.
We predict that the steady-state output photon rate 
can reach $\Phi_{\rm{out}}^{\rm{ss}}\sim 4\times 10^{-3}\omega$ in the NP, then suddenly
vanishes when tuning the Rabi Hamiltonian into the SP.
This demonstrates a sudden change of the ground eigenstate of the quantum Rabi Hamiltonian,
and indicates the occurrence of the superradiant QPT.

\section*{\large{Results}}

\section{\small {Superadiant quantum phase transitions}}
Note that the theory of the superradiant phase transition in the quantum Rabi model was
	developed first in 2013 \cite{Ashhab2013PRA} and then in
	2015 \cite{Hwang2015PRL}. For clarity, we first review the theory of the superradiant phase transition in the quantum Rabi model.
In the limit of $\Omega/\omega\rightarrow \infty$, following Refs.~\cite{Ashhab2013PRA,Hwang2015PRL}, we can solve
the Hamiltonian $H_{R}$ in Eq.~(\ref{eq1}) using
a Schrieffer-Wolff transformation \cite{Scully1997Book,Agarwal2012Book}: 
\begin{align}
	U_{\rm{SW}}=\exp{\left[\left(\frac{g}{\Omega}\right)\left(a+a^{\dag}\right)\left(|e\rangle\langle g|-|g\rangle\langle e|\right)\right]}.
\end{align}
For the transformed Hamiltonian, the transitions between the ground- and excited-qubit-state subspaces $\left\{|n\rangle|g\rangle\right\}$ and $\left\{|n\rangle|e\rangle\right\}$ are negligible. Thus, we perform a projection $\langle g|U_{\rm SW}^{\dag}H_{R}U_{\rm SW}|g\rangle$, resulting in
\begin{align}\label{eq2}
	H_{\rm{np}}=\omega a^{\dag}a-\frac{\omega g_{c}^{2}}{4}\left(a+a^{\dag}\right)^{2}+O(g_{c}^{2}\omega/\Omega),
\end{align} 
where $$g_{c}=2g/\sqrt{\omega \Omega},$$ is the normalized coupling strength and $O(g_{c}^{2}\omega/\Omega)$
denotes negligible higher-order terms. 
{The excitation energy of $H_{\rm np}$ is $$\varepsilon_{\rm{np}}=\omega\sqrt{1-g_{c}^{2}},$$
	which is real only for $g_{c}\leq 1$ and vanishes at $g_{c}=1$, i.e., in the NP \cite{Hwang2015PRL}. The ground eigenstate of $H_{\rm{np}}$ is $|E_{0}\rangle=S(r_{\rm{np}})|0\rangle$, with
\begin{align}
	S(r_{\rm{np}})=\exp\left[\frac{r_{\rm{np}}}{2}\left(a^{\dag 2}-a^{2}\right)\right],\ \ \ \ \ \  {\rm and}\ \ \ \ \ \ \
	r_{\rm{np}}=-\frac{1}{4}\ln\left(1-g_{c}^{2}\right).
\end{align}}

\begin{figure}
	\centering
	\scalebox{0.8}{\includegraphics{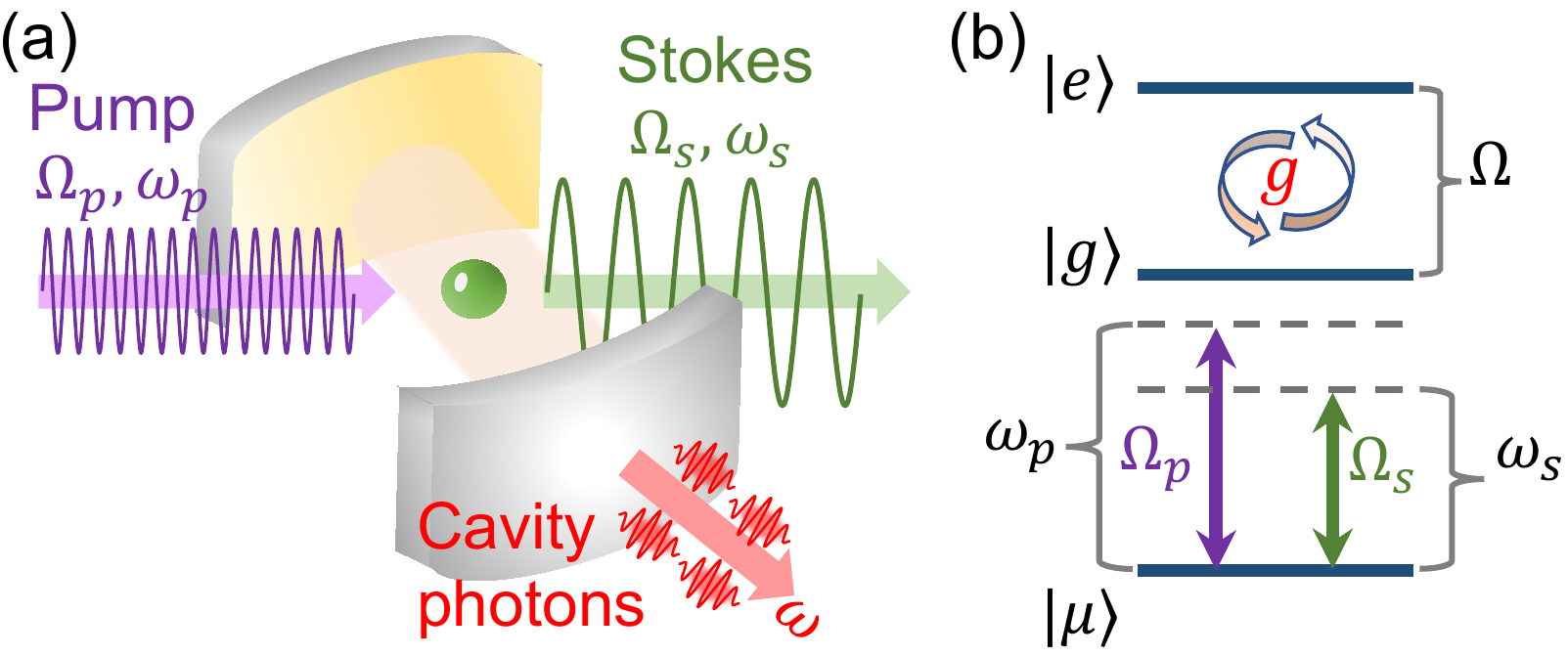}}
	\caption{\textbf{Sketch and atomic level structure of the protocol}. (a) Sketch of the parametric down-conversion process in an atom-cavity system. A pump pulse of frequency $\omega_{p}$ is converted into 
		a Stokes pulse of frequency $\omega_{s}$ and multiple cavity photons of frequency $\omega$. 
		(b) Level structure of the atom. The upper
		two atomic levels $|g\rangle$ and $|e\rangle$ are ultrastrongly coupled to the
		cavity mode with strength $g$. The lower two levels $|g\rangle$ and $|\mu\rangle$ are off-resonantly
		driven by a composite pulse of two frequencies ($\omega_{p}$ and $\omega_{s}$) and two amplitudes ($\Omega_{p}$ and $\Omega_{s}$).
	}
	\label{fig1}
\end{figure}

For $g_{c}>1$ and 
\begin{align}
  \alpha=\frac{1}{2}\sqrt{\frac{\Omega}{\omega}\left(g_{c}^{2}-g_{c}^{-2}\right)} \ \geq0,
\end{align}
after displacing the cavity field with displacement operators 
\begin{align}
	D(\pm\alpha)=\exp\left[\pm\alpha\left(a^{\dag}-a\right)\right],
\end{align}  
we can employ the same procedure in deriving $H_{\rm{np}}$ to obtain the Hamiltonian in the SP:
\begin{align}
	H_{\rm{sp}}\approx\omega a^{\dag}a-\frac{\omega}{4 g_{c}^{4}}\left(a+a^{\dag}\right)^{2}+{\frac{\Omega}{2}\left(1-g_{c}^{2}\right)+\omega\alpha^2}.
\end{align}
The excitation energy of $H_{\rm{sp}}$ is $\varepsilon_{\rm{sp}}=\omega\sqrt{1-g_{c}^{-4}}$ and 
the ground eigenstate of $H_{\rm{sp}}$ is $S(r_{\rm{sp}})|0\rangle$, with 
\begin{align}
	r_{\rm{sp}}=-\frac{1}{4}\ln\left(1-g_{c}^{-4}\right).
\end{align}
Thus, in the lab frame, $H_{R}$ has two degenerate ground eigenstates 
\begin{align}
	|E_{0}\rangle= D(\pm\alpha)S(r_{\rm{sp}})|0\rangle|\!\downarrow\rangle_{\pm},
\end{align}
where $|\!\downarrow\rangle_{\pm}$ are the ground eigenstates of 
\begin{align}
	H_{\pm}=\Omega|e\rangle\langle e|\mp2g\alpha(|e\rangle\langle g|+|g\rangle\langle e|).
\end{align}
This is because two different displacement parameters $\pm\alpha$ result in an indentical spectrum \cite{Ashhab2013PRA}.

\section{\small{Demonstrating the critical phenomenon}}
{The sudden change of $\bar{n}_{0}=\langle E_{0}|a^{\dag}a|E_{0}\rangle$ at the critical point $g_{c}=1$ is the most important benchmark of the superradiant QPT. Specifically, when $\Omega/\omega\rightarrow\infty$, the derivative $d\bar{n}_{0}/dg_{c}$ is discontinuous at the critical point $g_{c}=1$ [see Fig.~2(a)], indicating the occurrence of the superradiant QPT.
	However, observing this critical phenomenon is still challenging in experiments   
	mainly because: (i) it is difficult to prepare the ground state $|E_{0}\rangle$ at the critical point; (ii) one cannot adiabatically tune control
	parameters across the critical point \cite{Cai2021NC,Chen2021NC} since the energy gap mostly closes at the
	critical point; and (iii) the photons in the ground eigenstate $|E_{0}\rangle$ are virtual and cannot be directly measured \cite{Forn2019RMP,Kockum2019NRP}.}

\begin{figure}
	\centering
	\scalebox{0.45}{\includegraphics{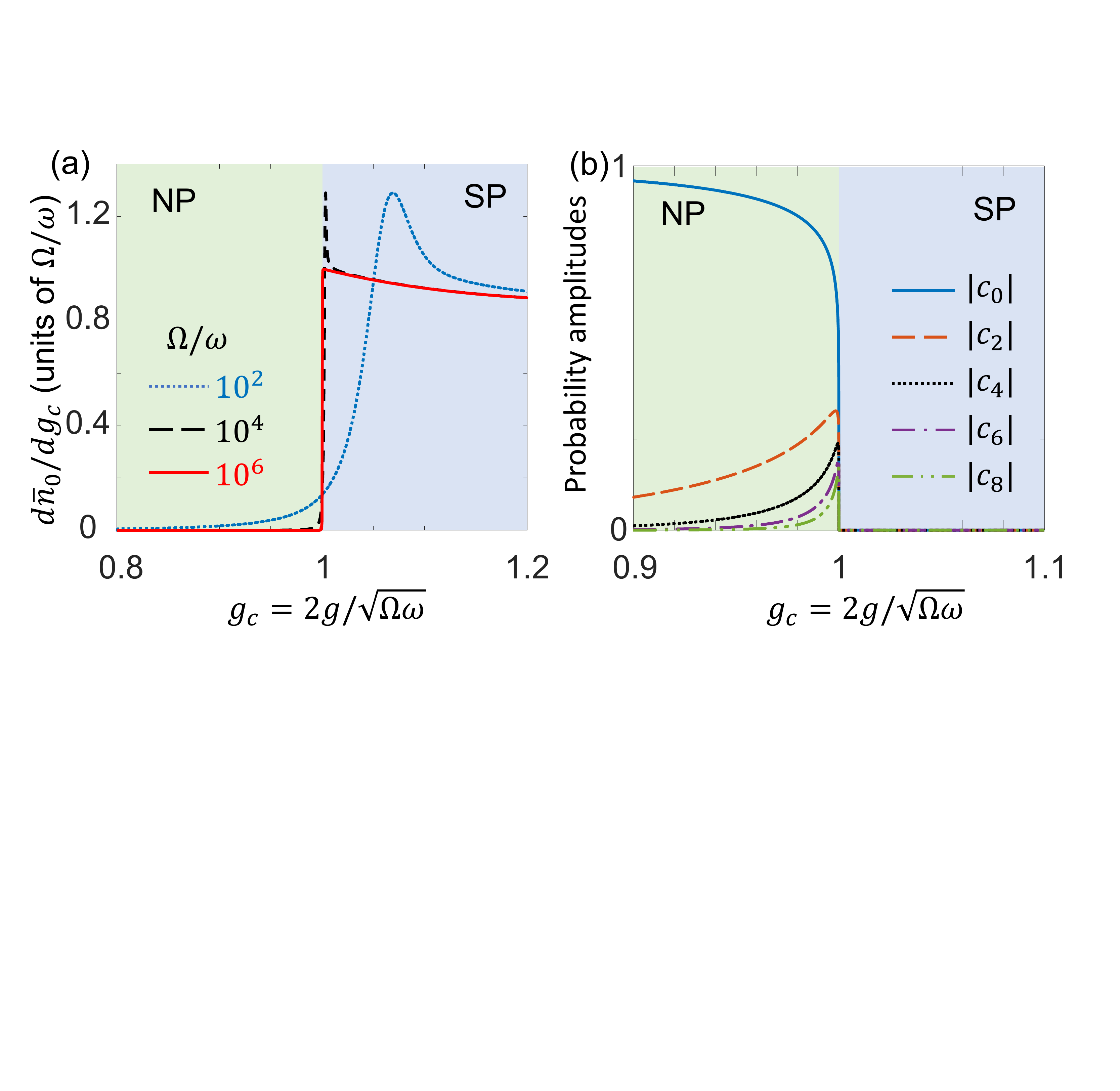}}
	\caption{\textbf{Sudden change of photon number distribution}. (a) Derivative $d\bar{n}_{0}/d g_{c}$ of the virtual cavity excitation number $\bar{n}_{0}=\langle E_{0}|a^{\dag}a|E_{0}\rangle$ calculated for different frequency rate $\Omega/\omega$. (b) Probability amplitudes $|c_{2k}|$ of the states $|2k\rangle|g\rangle$ in the eigenstate $|E_{0}\rangle$ of $H_{R}$ in Eq.~(\ref{eq1}) when $\Omega=10^{6}\omega$ approximatively satisfying the condition $\Omega/\omega\rightarrow \infty$. The green- and blue-shaded areas denote the normal phase (NP) and the superradiant phase (SP), respectively.
	}
	\label{fig2}
\end{figure}

{Instead, in this manuscript, we propose to measure the sudden change of the photon number 
	distributions of $|E_{0}\rangle$, i.e., the probability amplitudes $c_{k}(g_{c})=\langle g|\langle k|E_{0}\rangle$.
	This is equivalent to measuring the change of $\langle E_{0}|a^{\dag}a|E_{0}\rangle$ because it can be calculated using the photon number 
	distributions.}
In the $\Omega/\omega\rightarrow \infty$ limit, it is expected that $\alpha\rightarrow \infty$ and $r_{\rm{sp}}\neq0$ when $g_{c}>1$.
{Thus, when $g_{c}$ is tuned across the critical point, there is a sudden change in the amplitude $c_{2k}$:
\begin{align}
&|c_{2k}(g_{c})|\xrightarrow{\ g_{c}=1^{-}}\left|\frac{\left[-\tanh(r_{\rm np})\right]^{k}\sqrt{(2k)!}}{2^{k}k!\sqrt{\cosh(r_{\rm np})}}\right|>0,\cr
&|c_{2k}(g_{c})|\xrightarrow{\ g_{c}=1^{+}}\left|\frac{(\tanh r_{\rm sp})^{k}\exp\left[-\frac{\alpha^2}{2}(1+\tanh r_{\rm sp})\right]}{2^{k}\sqrt{(2k)!\cosh r_{\rm sp}}}\mathcal{H}_{_{n}}(x)\right|\longrightarrow 0,
\end{align}
where $\mathcal{H}_{n}(x)$ are the Hermite polynomials with 
\begin{align}
	x=\frac{\alpha \exp{(r_{\rm sp})}}{\sqrt{\sinh2r_{\rm sp}}}.
\end{align}
Such a change is obvious when $k$ is small, because $|c_{2k}(g_{c}\rightarrow 1^{-})|$ is significant as shown in Fig.~2(b).
Especially, for $k<5$, 
the components of the few-photon states in the eigenstate $|E_{0}\rangle$ of $H_{R}$ suddenly vanish when $g_{c}$ is tuned across the
critical point [see Fig.~2(b)].} {This coincides with the sudden changes of the photon number $\bar{n}_{0}$ and its derivative $d\bar{n}_{0}/dg_{c}$ [see the red-solid curve in Fig.~2(a)].} 

To demonstrate the sudden change of $c_{2k}$, {following the idea in Ref.~\cite{Huang2014PRA}}, we introduce a third low-energy level $|\mu\rangle$ with level frequency $\omega_{\mu}<0$ for the atom [see Fig.~1(b)].
{We assume that $|\omega_{\mu}|\gg\omega$, so that the state $|\mu\rangle$ is far off-resonance to
	the cavity bare frequency.}
The atom-cavity interaction becomes 
\begin{align}
	H_{0}=H_{R}+\omega_{\mu}| \mu\rangle\langle \mu|+\omega a^{\dag}a|\mu\rangle\langle\mu|,
\end{align}
whose eigenstates $|\xi_{j}\rangle$ can be separated into: (i) the noninteracting sectors $|\mu_{n}\rangle=|n\rangle|\mu\rangle$ with eigenvalues $n\omega+\omega_{\mu}$;
and (ii) the eigenstates $|E_{m}\rangle$ of $H_{R}$ with eigenvalues $E_{m}$ ($j,n,m=0,1,2,\ldots$).
The ground state of the whole system becomes $|\mu_{0}\rangle$, 
which is the initial state for the system dynamics hereafter.

\begin{figure}
	\centering
	\scalebox{0.7}{\includegraphics{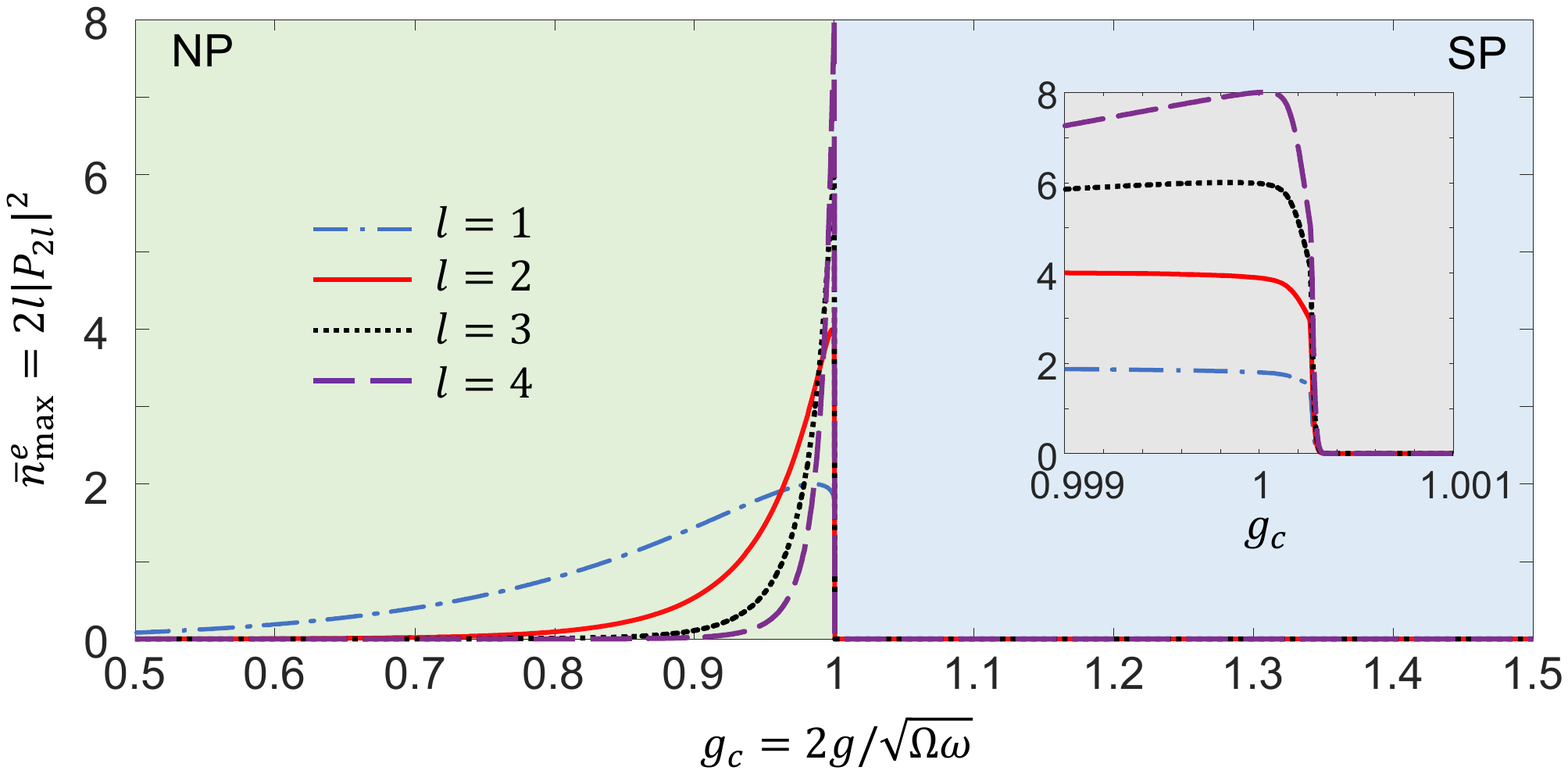}}
	\caption{\textbf{Critical phenomenon}. Theoretical prediction of the maximum photon number of the system after a finite-time evolution, according to Eqs. (\ref{eq7}) and (\ref{eq17a}). For $\Xi\rightarrow 0$, we impose $t=10^6/\omega$ when $\pi/\Xi\rightarrow\infty$, to avoid an infinite evolution time in the coherent dynamics. Parameters are: $\Omega=10^{6}\omega$, $\omega_{\mu}=E_{0}-\left[2(n_{d}-l)+0.25\right]\omega$, $\Omega_{p}=0.005(E_{2}-E_{0})$, and $\Omega_{s}=2\Omega_{p}$. For simplicity, we choose $n_d = 2+l$ in the simulation.
	}
	\label{fig3}
\end{figure}

Then, as shown in Fig.~1, we employ a composite pulse to drive the atomic transition $|\mu\rangle\leftrightarrow|g\rangle$:
\begin{align}
	H_{D}=\left[\Omega_{p}\cos(\omega_{p}t)+\Omega_{s}\cos(\omega_{s}t)\right]\left(|\mu\rangle\langle g|+|g\rangle\langle\mu|\right),
\end{align} 
where $\Omega_{p,(s)}$ and $\omega_{p,(s)}$ are the driving amplitude and frequency, respectively.
Focusing on the changes of few-photon components ($k<5$), we choose 
$\omega_{p}=E_{0}-\omega_{\mu}$ and $\omega_{s}=E_{0}-\omega_{\mu}-2l\omega$, for $l=1,2,\ldots$,
resulting in the resonant transitions $|\mu_{0}\rangle\leftrightarrow|E_{0}\rangle\leftrightarrow|\mu_{2l}\rangle$ in the
subspace $\{|\xi_{j}\rangle\}$.
Specifically, for weak drivings $\Omega_{p,(s)}\ll |E_{2}-E_{0}|$, 
the system dynamics is understood as the Raman transitions
described by the effective Hamiltonian 
\begin{align}\label{eq5}
	H_{\rm{eff}}=\frac{1}{2}\left[c_{0}\Omega_{p}|\mu_{0}\rangle+c_{2l}\Omega_{s}|\mu_{2l}\rangle\right]\langle E_{0}|+{\rm{h.c.}},
\end{align}
which is obtained by performing $e^{iH_{0}t}H_{D}e^{-iH_{0}t}$ and neglecting fast-oscillating terms under the rotating-wave
approximation (see Supplementary
Note 1 for more details). {Hereafter,
we omit the explicit $g_{c}$
dependence of all the probability amplitudes $c_{k}$ for simplicity.}
The dynamics governed by $H_{\rm{eff}}$ can be interpreted as a multi-photon down-conversion process [see Fig.~1(a)],
with an  
efficiency controllable by the coefficients $c_{0}$ and $c_{2l}$. 
When the initial state is $|\mu_{0}\rangle$, the probability amplitude of the state $|\mu_{2l}\rangle$ at time $t$ reads
\begin{align}\label{eq7}
	P_{2l}=\frac{c_{0}\Omega_{p} c_{2l}\Omega_{s}}{4\Xi^2}\left[\cos{\left(\Xi t \right)-1}\right], \ \ \ \ {\rm with}\ \ \ \ 
	\Xi^2=\frac{1}{4}\left[\left(c_{0}\Omega_{p}\right)^{2}+\left(c_{2l}\Omega_{s}\right)^{2}\right].
\end{align}
Choosing $t=\pi/\Xi$, the probability amplitude$P_{2l}$ and the mean photon number of the system both reach their maximum values. Then, as long as $\Omega_{p}\neq0$ and $\Omega_{s}\neq 0$, $P_{2l}$ is measurable because $c_{0}$ and $c_{2l}$ are significant. 
Accordingly, the theoretical prediction of the maximum photon number of the system after a finite-time evolution is
\begin{align}\label{eq17a}
	\bar{n}_{\rm max}^{e}=2l|P_{2l}(t)|^{2}=2l\left|\frac{2c_{0}\Omega_{p}c_{2l}\Omega_{s}}{\left(c_{0}\Omega_{p}\right)^2+\left(c_{2l}\Omega_{s}\right)^2}\right|^2. \ \ \ \ \ \ \ \ \ \ \ ({\rm where}\ t=\pi/\Xi)
\end{align}	

However, when $g_{c}$ is tuned across the critical point, the needed evolution time to achieve the
maximum photon number tends infinite due to $\Xi\rightarrow 0$.
To avoid such an infinite-time evolution, we impose $t\leq 10^{6}/\omega$ in this protocol when $\Xi\rightarrow 0$. In this limit, we obtain $P_{2l}\rightarrow 0$ because $\cos(\Xi t)\rightarrow 1$ according to Eq.~(\ref{eq7}). Therefore, as shown in Fig.~3,
we theoretically predict a sudden change of the mean photon number when $g_{c}\rightarrow 1$, 
indicating the occurrence of the superradiant QPT.


\begin{figure}
	\centering
	\scalebox{0.5}{\includegraphics{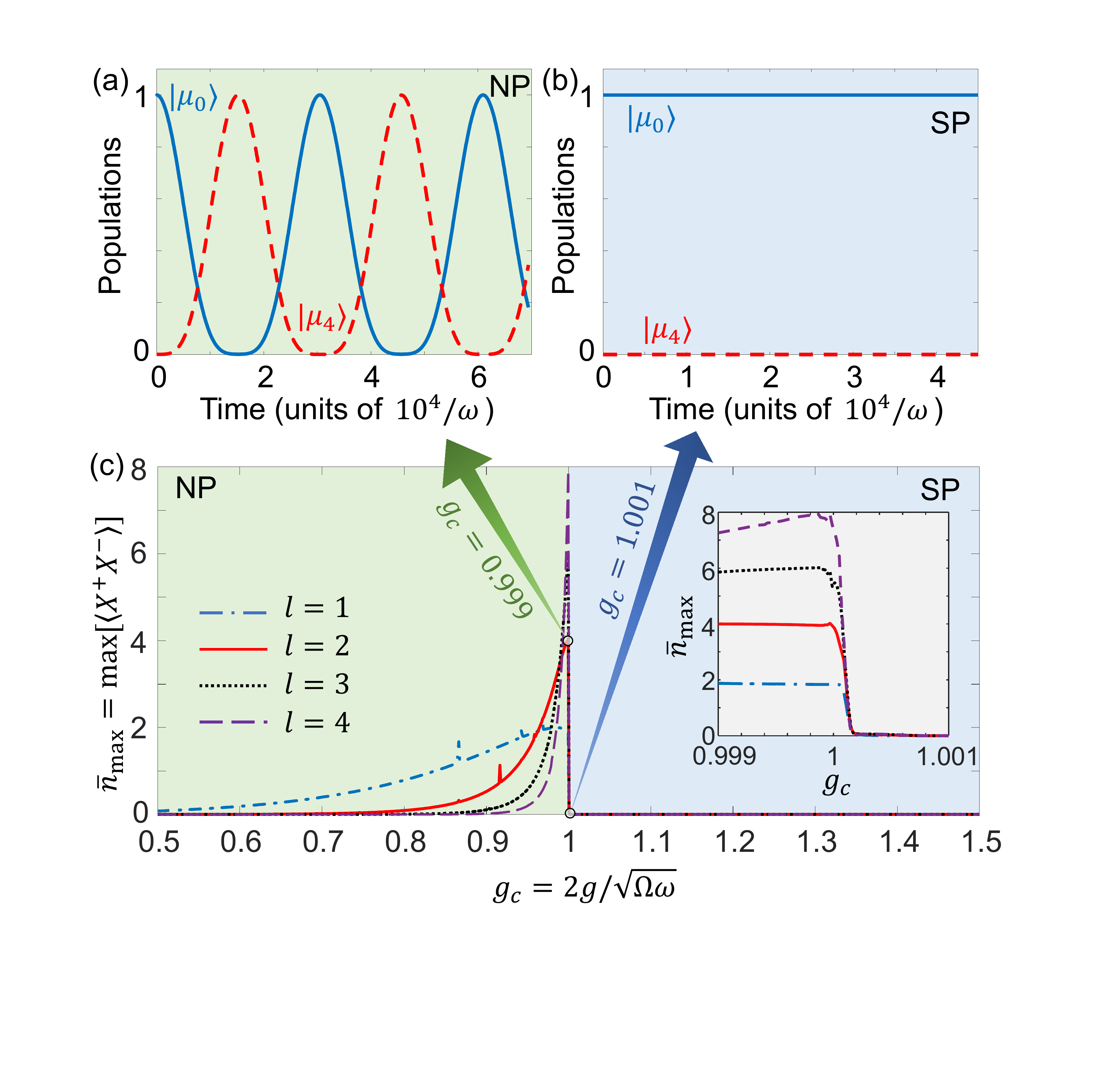}}
	\caption{\textbf{Dynamics of the model}. Populations of the ground state $|\mu_{0}\rangle$ and the four-photon state $|\mu_{4}\rangle$ in (a) the normal phase (NP, green-shaded area) for $g_{c}=0.999$, and (b) the superradiant phase (SP, blue-shaded area) for $g_{c}=1.001$, calculated for the total Hamiltonian $H_{\rm{tot}}=H_{0}+H_{D}$. (c) Maximum value ($\bar{n}_{\rm{max}}$ at the time $t=\pi/\Xi$) over time of the mean photon number in the evolution governed by $H_{\rm{tot}}$ for $l=1,2,3,4$. Parameters are the same as those in Fig. 3. Here, $n_{d}\geq l$ is used to tune the driving frequencies $\omega_{p}$ and $\omega_{s}$; {and for simplicity, we choose $n_d=2+l$ in the simulation.} The eigenvalues $E_{0}$ and $E_{2}$ can be numerically calculated. The level frequency $\omega_{\mu}$ is chosen so that the state $|E_{0}\rangle$ is the highest level in the dynamical evolution and the state $|\mu\rangle$ is off-resonant to the cavity bare frequency. Thus, the system only has real cavity photons contributed by the state $|\mu_{2l}\rangle$. 
	}
	\label{fig4}
\end{figure}


This demonstration can be
interpreted as
a partial quantum process tomography,
which starts at preparing the initial state $|\mu_{0}\rangle$ and fixing $g_{c}$ to a specific value, e.g., $g_{c}=0.999$. Then, turning on the drivings, one can detect the system dynamics [see the example in Fig.~4(a)].
After the detection, the drives are turned off and the system is prepared to the state $|\mu_{0}\rangle$. {The next step is tuning $g_{c}$ to another value (e.g., $g_{c}=1.001$) through adjusting the frequencies $\Omega$ and $\omega$}. Thus, after the
energy spectrum of the system stabilizes,
one can turn on the drivings again and detect the system dynamics [see the example in Fig.~4(b)].
By repeating these processes, we can know the system dynamics for an arbitrary $g_{c}$
and demonstrate the critical phenomenon as shown in Fig.~4(c), which shows the maximum value $\bar{n}_{\rm{max}}$ of the mean photon number of the system in the evolution.
Note that the mean photon number is no longer $\bar{n}=\langle a^{\dag}a\rangle$, but
$\bar{n}=\langle X^{-}X^{+}\rangle$, where $X^{+}$ ($X^{-}$) is the positive (negative) frequency component of the quadrature operator $X=a+a^{\dag}$ \cite{Ciuti2006PRA,Ridolfo2012PRL,Kockum2019NRP,Forn2019RMP}.
Otherwise, because of $\langle E_{0}|a^{\dag}a|E_{0}\rangle\neq 0$, an observation of the stream of photons in the eigenstate $|E_{0}\rangle$ of the Rabi model would give rise to a perpetuum mobile
behavior \cite{Ciuti2006PRA,Stassi2013PRL}. 
Because these photons can be emitted and detected in a dissipative dynamics, a measurement of the population dynamics is not necessary.

Using the corresponding input-output theory \cite{Ridolfo2012PRL,Huang2014PRA}, we apply 
\begin{align}\label{eq17}
  X^{+}=\sum_{j,j'}\langle \xi_{j'}|\left(a+a^{\dag}\right)|\xi_{j}\rangle|\xi_{j'}\rangle\langle \xi_{j}|,
\end{align} 
and $X^{-}=\left(X^{+}\right)^{\dag}$ with $j'<j$. Here, Eq.~(\ref{eq17}) describes that 
a photon emission from the cavity is associated with the transition
from a high-energy eigenstate $|\xi_{j}\rangle$ to a low-energy eigenstate $|\xi_{j'}\rangle$ of $H_{0}$.
Note that in the subspace $\{|\mu_{n}\rangle\}$, we can obtain $\sum_{n}|\mu_{n}\rangle\langle\mu_{n}|X^{+}\sum_{m}|\mu_{m}\rangle\langle\mu_{m}|=|\mu\rangle\langle\mu|\otimes a$. That is, the photons in the states $|\mu_{2l}\rangle$ are real cavity photons, thus, 
\begin{align}
	X^{-}X^{+}|\mu_{2l}\rangle\equiv |\mu\rangle\langle\mu|\otimes a^{\dag}a|\mu_{2l}\rangle=a^{\dag}a|2l\rangle|\mu\rangle.
\end{align}
Therefore, the system has maximum real photons at the time $t=\pi/\Xi$, when the state $|\mu_{2l}\rangle$ is maximally populated according to Eq.~(\ref{eq7}).
Figure 4(c) shows that $\bar{n}_{\rm{max}}$ reaches a maximum when $g_{c}\rightarrow 1^{-}$,
indicating a rapid increase of the mean photon number near the critical point.
Then, when $g_{c}$ is tuned across the critical point, the photons suddenly
vanish. 
The inset of Fig.~4 shows such changes more clearly, where we can see that $\bar{n}_{\rm{max}}$
changes sharply when slightly increasing $g_{c}$ from $g_{c}= 1$ to $g_{c}=(1+10^{-4})$, demonstrating a 
sudden change of the photon-number distributions in $|E_{0}\rangle$ (see also Supplementary
Note 1).
Obviously, the actual dynamics of the system shown in Fig.~4(c) coincides very well with the effective one shown in Fig.~3.
As an example for $l=2$, when $g_c=0.99999$, the theoretical prediction 
	of the maximum photon number
	is $\bar{n}_{\rm max}^{e}\simeq 3.89$, and the actual number is $\bar{n}_{\rm max}\simeq 3.93$.
	When we change $g_{c}$ to $g_{c}=1.0003$, the theoretical prediction becomes $\bar{n}_{\rm max}^{e}\simeq 0.0018$ and the actual number is $\bar{n}_{\rm max}\simeq 0.04$. Both close to zero,
indicating a sudden change of the mean photon number at the critical point.
The sudden change in $\bar{n}_{\rm{max}}$ can be easily detected experimentally by measuring 
the rate of photons transmitted out of the cavity.
Note that numerical results for actual dynamical evolution in this manuscript are obtained using
the total Hamiltonian $H_{\rm{tot}}=H_{0}+H_{D}$ without approximations.


\section{\small{Output photon rate}}
A proper
generalized input-output relation in the ultrastrong-coupling
regime determines the output cavity photon rate \cite{Stassi2013PRL,Huang2014PRA} by 
\begin{align}\label{eq8}
	\Phi_{\rm{out}}=\kappa {\rm{Tr}}\left[X^{-}X^{+}\rho\right].
\end{align}
Here, $\kappa$ is the cavity decay rate and $\rho$ is the density matrix obeying
the following master equation at zero temperature \cite{Kockum2019NRP,Forn2019RMP},
\begin{align}\label{eq9}
	\dot{\rho}=i\left[\rho,H_{\rm{tot}}\right]+\sum_{\nu=1}^{3}\sum_{j,j'<j}\Gamma_{jj'}^{(\nu)}\left\{\mathcal{D}\left[|\xi_{j'}\rangle\langle\xi_{j}|\right]\rho\right\},
\end{align}
where $H_{\rm{tot}}=H_{0}+H_{D}$ is the total Hamiltonian and 
\begin{align}
	\mathcal{D}[o]\rho=o\rho o^{\dag}-\frac{1}{2}\left(o^{\dag}o\rho+\rho o^{\dag}o\right),
\end{align}
is the Lindblad superoperator.
The relaxation coefficients $\Gamma_{jj'}^{(\nu)}$ can be written in compact forms
as 
\begin{align}
	\Gamma_{jj'}^{(1)}=&\gamma_{1}|\langle\xi_{j'}|(|\mu \rangle\langle g|+|g \rangle\langle \mu|)|\xi_{j}\rangle|^{2},\cr
	\Gamma_{jj'}^{(2)}=&\gamma_{2}|\langle\xi_{j'}|(|g\rangle\langle e|+|e \rangle\langle g|)|\xi_{j}\rangle|^{2}, \cr
	\Gamma_{jj'}^{(3)}=&\kappa|\langle\xi_{j'}|(a+a^{\dag})|\xi_{j}\rangle|^{2}, 	
\end{align}
where $\gamma_{1,(2)}$ is the spontaneous emission rate of the
transition $|g\rangle\rightarrow|\mu\rangle$ ($|e\rangle\rightarrow|g\rangle$).

\begin{figure}
	\centering
	\scalebox{0.5}{\includegraphics{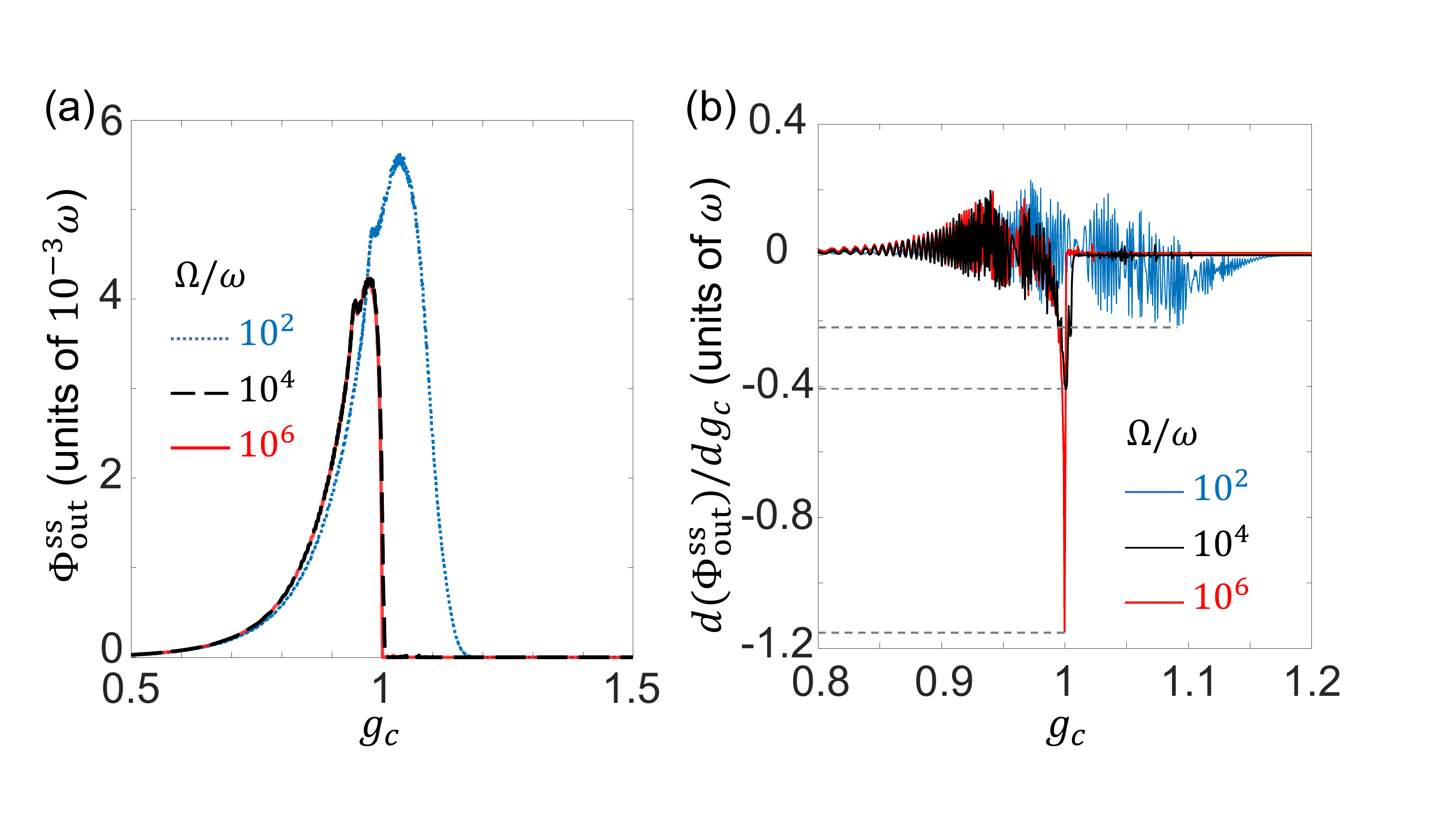}}
	\caption{\textbf{Photon output field}. (a) Steady-state output photon rates $\Phi_{\rm{out}}^{\rm{ss}}=\Phi_{\rm{out}}|_{t\rightarrow \infty}$ defined in Eq.~(\ref{eq8}), and (b) the corresponding derivative $d\left(\Phi_{\rm{out}}^{\rm{ss}}\right)/dg_{c}$ versus $g_{c}$ for $l=2$. Dissipation rates are $\kappa=\gamma_{1}=\gamma_{2}=0.01\omega$. We choose relatively strong driving fields, i.e., $\Omega_{p}=0.05(E_{2}-E_{0})$ and $\Omega_{s}=2\Omega_{p}$, to achieve relatively large output photon rates. Strong driving fields may cause small errors (via counter-rotating effects) in obtaining the effective Hamiltonian $H_{\rm{eff}}$, leading to oscillations in $\Phi_{\rm{out}}^{\rm{ss}}$. These small errors do not affect the observation of the critical phenomenon, i.e., an extremum of $d\left(\Phi_{\rm{out}}^{\rm{ss}}\right)/dg_{c}$ at $g_{c}\rightarrow 1^{+}$. The blue, black, and red curves are plotted using frequency ratios $\Omega/\omega=10^{2}$, $\Omega/\omega=10^{4}$, and $\Omega/\omega=10^{6}$, respectively. Other parameters are the same as those in Fig.~4.
	}
	\label{fig5}
\end{figure}

The steady-state output field can remain unchanged for a long time.
	This can reduce the difficulties in experiments to measure the emitted photons. Taking $l=2$ as an example, in Fig.~5(a), we show the steady-state output photon rates $\Phi_{\rm{out}}^{\rm{ss}}=\Phi_{\rm{out}}|_{t\rightarrow\infty}$, which is independent of the initial state.
Focusing on $\Omega/\omega=10^{4}$ [the black-solid curve in Fig.~5(a)],
the peak value of steady-state output photon rate can reach $\Phi_{\rm{out}}^{\rm{ss}}\gtrsim 4\times10^{-3}\omega$ in the NP near the critical point when the dissipation 
rates are $\kappa=\gamma_{1}=\gamma_{2}=0.01\omega$. We can choose $\omega\sim 2\pi\times 5~$MHz and $\Omega\sim 2\pi\times 50$~GHz \cite{Yoshihara2016NP,ChenPRA2017},
the cavity can transmit $\sim 1.25\times 10^{5}$ photons per second, which is detectable in cavity- and circuit-QED systems.
The coupling strength is $g\sim 2\pi\times 250$~MHz, {and the driving amplitudes are $\Omega_{p}\gtrsim 2\pi\times 0.5$~kHz and $\Omega_{s}\gtrsim2\pi\times1$~kHz, which
	are feasible in current experiments \cite{Gu2017PR,Forn2019RMP,Kockum2019NRP,Krantz2019APR,Kjaergaard2020Arc,Kwon2021Jap}.}
When the parameter $g_{c}$ crosses the critical point, suddenly there is no photon released from the cavity.
This indicates the drastic change of the photon number distributions and the occurrence of the QPT.
The curves in Fig.~5(a) coincide well with those in Fig.~2(a),
proving that the steady-state output field is a good signature of the superradiant QPT.

\section{\small{Finite-frequency effect}}
As mentioned above, for finite frequencies of $\Omega$ and $\omega$,
the higher-order terms $O(\omega g_{c}^{2}/\Omega)$ cannot be ideally neglected but modify the
eigenvalues and eigenstates of $H_{R}$ near the critical point. 
The influence of this finite-frequency effect is shown in Fig.~5.
The first-order derivatives of $\Phi_{\rm{out}}^{\rm{ss}}$ versus $g_{c}$
are shown in Fig.~5(b).
For $\Omega/\omega= 10^4$ [black-solid curve in Fig.~5(b)] and  $\Omega/\omega= 10^6$ [red-solid curve in Fig.~5(b)], we can see deep thin dips near the critical point of $g_{c}=1$,
indicating the sudden changes of output photon rates. 
For $\Omega/\omega=10^{2}$, the phenomenon becomes less pronounced [see the blue-solid curve in Fig.~5(b)].

\section*{\large{Discussion}}
Our protocol does not need to control a parameter to adiabatically approach the critical point. 
Also, we do not need to prepare the equilibrium state, which is challenging in experiments because the time required diverges \cite{Hwang2015PRL,Garbe2020PRL}.
The critical phenomenon can be translated to a sudden change of the output photon rate, which can be easily detected in experiments.
Thus, our protocol could be efficient to observe the critical phenomenon of the sudden change of photon number distributions in a superradiant QPT.
Moreover, unlike dissipative phase transitions \cite{Baumann2010Nat,Zhang2017Opt,Klinder2015PNAS,Lonard2017Nat,Minganti2021NJP,Minganti2021PRR}, the drivings and dissipation in this manuscript are only used for quantum measurements and do not affect the QPT.

Our protocol could be implemented with superconducting
circuits, which have realized the ultra- and deep-strong couplings between a single atom and a single-mode cavity \cite{Niemczyk2010NP,Forn2010PRL,Yoshihara2016NP,Yoshihara2017PRA,ChenPRA2017,Bosman2017Njpqi,Yoshihara2018PRL}. One can
couple, e.g., a flux qubit \cite{Gu2017PR,Forn2019RMP,Kockum2019NRP,Krantz2019APR,Kjaergaard2020Arc,Kwon2021Jap}, with a lumped-element resonator via Josephson
junctions to reach a coupling strength of $g/2\pi\sim 1$~GHz \cite{Yoshihara2016NP,ChenPRA2017}.
To reduce the cavity frequency to $\omega=2\pi\times 5$~MHz,
one can use an array of dc superconducting quantum interference devices \cite{Liao2010PRA} instead of the lumped-element resonator (see Supplementary
Note 2). 
The level
structure of the atom can be realized by adjusting the external magnetic flux through the qubit loop \cite{Stassi2013PRL,Huang2014PRA,Stefano2017NJP}.
Moreover, simulated quantum Rabi models \cite{Forn2019RMP,Kockum2019NRP,Ballester2012PRX,Lv2018PRX,Qin2018PRL,Leroux2018PRL,Carlos2020PRA,Chen2021PRL}, which work in the rotating frames of the Jaynes-Cummings models, 
can be another possible implementation of our protocol (see also Supplementary
Note 3).

\section*{\large{Conclusion}}
We have investigated a method to observe quantum critical phenomena: 
a sudden change of photon number distributions in a quantum phase transition exhibited by the quantum Rabi model. 
We can interpret the system dynamics as a multi-photon down-conversion process and study the outputs near the critical point.
When the normalized coupling strength $g_{c}$ is tuned across the critical point,
the output of the system changes abruptly.
This reflects a sudden change of the eigenstate $|E_{0}\rangle$ of the quantum Rabi model. 
Specifically, for the Rabi Hamiltonian in the NP, a pump pulse can be converted into a Stokes pulse and multiple cavity photons, while in the SP, it cannot.
One can observe such a change by measuring the photons emitted out
of the cavity continuously in the steady state.
Moreover, for finite frequencies, we demonstrate that a small frequency ratio $\Omega/\omega$ may lead to
the disappearance of the critical phenomenon. 
We expect that our method can provide a useful tool to study various critical phenomena
exhibited by quantum phase transitions without preparing their equilibrium states.

\section*{\large{Acknowledgements}}
Y.-H.C. was supported by the National Natural Science Foundation of China
under Grant No. 12304390.
A.M. was supported by the Polish National Science Centre (NCN) under the Maestro Grant No. DEC-2019/34/A/ST2/00081.
W.Q. was supported in part by the Incentive Research Project of RIKEN.
Y.X. was supported
by the National Natural Science Foundation of China
under Grant No. 11575045, the Natural Science Funds
for Distinguished Young Scholar of Fujian Province under
Grant 2020J06011 and Project from Fuzhou University
under Grant JG202001-2.
F.N. is supported in part by: 
Nippon Telegraph and Telephone Corporation (NTT) Research, 
the Japan Science and Technology Agency (JST) 
[via the Quantum Leap Flagship Program (Q-LEAP), and the Moonshot R\&D Grant Number JPMJMS2061], 
the Asian Office of Aerospace Research and Development (AOARD) (via Grant No. FA2386-20-1-4069), and the Office of Naval
Research (ONR).

\section*{Author contributions}
Y.-H.C. conceived and developed the idea. Y.Q., A.M., N.L., and R.S. analyzed the data and
performed the numerical simulations, with help from S.-B.Z. and W.Q.. Y.-H.C., Y.X., and F.N cowrote the paper with
feedback from all authors.

\section*{Data availability}

The data used for obtaining the presented numerical results as well as for generating
the plots is available on request. Please refer to yehong.chen@fzu.edu.cn

\section*{Competing interests}

The authors declare that they have no competing interests.

\section*{\large{Reference}}

\bibliography{references}

\end{document}